# Dynamics of students' epistemological framing in group problem solving

Hai D. Nguyen[1], Deepa N. Chari[1], Eleanor C. Sayre[1*]

[1]*Department of Physics, Kansas State University, Manhattan, Kansas, USA*
[*]esayre@ksu.edu

**Abstract:** Many studies have investigated students' epistemological framing when solving physics problems. Framing supports students' problem solving as they decide what knowledge to employ and the necessary steps to solve the problem. Students may frame the same problem differently and take alternate paths to a correct solution. When students work in group settings, they share and discuss their framing to decide how to proceed in problem solving as a whole group. In this study, we investigate how groups of students negotiate their framing and frame shifts in group problem solving.

## I. Introduction

Physics problems at the upper-division level involve combining complex physics concepts and mathematical tools. Many studies have explored students' discourse during problem solving through an epistemic lens (Bing & Redish, 2009; Irving, Martinuk, & Sayre, 2013; S. F. Wolf, 2014). Some studies focus on contextual details to present epistemic frames (Bing & Redish, 2012), which depict how students understand a problem and how they decide which physics concepts and mathematical tools are appropriate in the given problem. Another perspective of epistemic framing focuses on the behaviors that students (or groups of students) exhibit when engaging with a context (Scherr & Hammer, 2009).

In real physics classrooms, students are often encouraged to solve problems in groups. In these situations students discuss, share, negotiate, or challenge each other's framing. A detailed investigation of the dynamics of epistemic framing within a group should reveal not only how students frame the context, but also how they interact with others and shape the group's framing. In this study, we are interested in understanding how students negotiate their epistemic frames and frame shifts in a group. We employed our two-axis framework for students' use of math in the context of physics (Modir, Thompson, & Sayre, submitted) to understand the dynamics of groups' epistemic framing.

## II. Background & literature review

In this section, we first present a brief review of some epistemic frame models to understand students' problem solving. Then we review research about group work and students' epistemic framing during group work. Finally we state our research question which emphasizes the focus of this paper.

Previous research on epistemic framing has identified a discrete number of frames that students may be engaged in at certain time. Bing and Redish introduced four epistemic frames to describe students' use of math in a physics context, namely *physical mapping, calculation, invoking authority, math consistency* (Bing & Redish, 2009). The *physical mapping* frame describes students' efforts in mapping physics ideas or understanding of a given physical system to mathematical constructs. In a *calculation* frame, students purely apply mathematical tools or follow computational procedures. In the *authority* frame, students invoke and trust information that comes from an authoritative source without any further justification. Students engage in the *math consistency* frame when they make sense of a new situation by relying on the similarity in mathematical structure between the new situation with a known situation. Among these four frames, the *physical mapping* and *calculation* frames are most closely related to students' application of math in a physics context, while the *math consistency* frame relates to how they check the validity of their mathematical reasoning. Under this model, students start problem solving with analyzing the physics scenario, then map it to a mathematical construct and further apply computational tools on that construct to obtain the final result. Thus, a problem requires students to shift from a physical mapping frame to the calculation frame to obtain a result. They may go back and forth between these epistemic frames when solving a problem. In terms of the "flow" of problem solving, this model is quite similar to the one proposed by the ACER framework (Wilcox et al, 2013), though the focus is on students' framing and not the steps of problem solving.

If a student arrives at an incorrect final result, she might revisit her calculations to find the mistake if she frames the problem solving activity in *calculation* frame. On the other hand, she might reconsider how she understands and maps the physical scenario to mathematics if she frames the problem solving activity in *physical mapping* frame (Bing & Redish, 2012). If the mistake lies in mapping the physics scenario to mathematics, but the student frames the activity as following a computational procedure, she may get stuck and might never attain the correct result. Framing allows researchers to seek causal explanations of difficulties in students' problem solving, unlike frameworks which model only the steps of problem solving (such as ACER). Later work on student framing using math and physics has explored the differences between the *conceptual physics*, *conceptual mathematics,* and *algorithmic mathematics* frames (Thompson, Modir, & Sayre, 2016).

Group work has been advocated as bringing several advantages to students' problem solving, especially in upper-division physics classes (Cerny, 2012). One of these advantages is that the "holes" in each student's knowledge can easily be made evident and filled with knowledge from other group members. In an epistemic framing perspective, discussion among group members may lead to a more productive frame which helps them solve problems more quickly and correctly.

Small group work analysis has proven to be useful in understanding students' framing (Bing & Redish, 2009; Cerny, 2012; Irving et al., 2013; S. F. Wolf, 2014; Scherr & Hammer, 2009). Scherr and Hammer describe how small group discussions during a tutorial can be intellectually demanding and intense. They relate such discussions to a way of exploring mechanistic reasoning of the physical phenomenon (Scherr & Hammer, 2009). As students share ideas together, their mechanistic reasoning becomes more detailed and correct. In the math-in-physics context, Bing and Redish examined students' tendencies to call upon group members to evaluate mathematical work during problem solving (Bing & Redish, 2009). They argue that group members' requests for justification present new bids in response to contested ideas. The discourse involves negotiations for the current epistemic frame and frame shifts until the group reaches consensus. Thus, group members' framing and negotiations cultivate group framing in a productive way.

Expanding Scherr and Hammer's (2009) discussion and joking frames, Irving, Martinuk, & Sayre (2013) developed two axes of student framing in upper-division problem solving in classical mechanics. Problem solving could be more expansive or narrow, and more serious or silly. They showed that students flowed among all four quadrants in their two-axis framework to productively solve problems. However, even though their students used mathematics extensively, they did not focus on mathematical nature of the students' problem solving. Modir, Thompson, & Sayre (2016) picks up the idea and development of the two-axis framework and applies it to the use of mathematics in upper-division quantum mechanics. We use that framework and apply it to student groups in E&M (detailed below).

In this study, our research question is: how do students negotiate a group's framing in problem solving? We are interested in how a group of students decides how to frame a problem solving activity and what frame shifts they make as they progress through in order to find a solution to the problem.

## III. Theoretical framework

We apply our two dimensional framework consisting of two orthogonal axes as illustrated in Figure 1 below. The horizontal axis describes students' activities towards the conceptual or algorithmic ends, while the vertical axis extends towards the physics and math ends. These orthogonal axes make up four quadrants, corresponding to four frames students may be in when solving physics problems involving math. These four frames and examples of students' activities in each frame are described as follows.

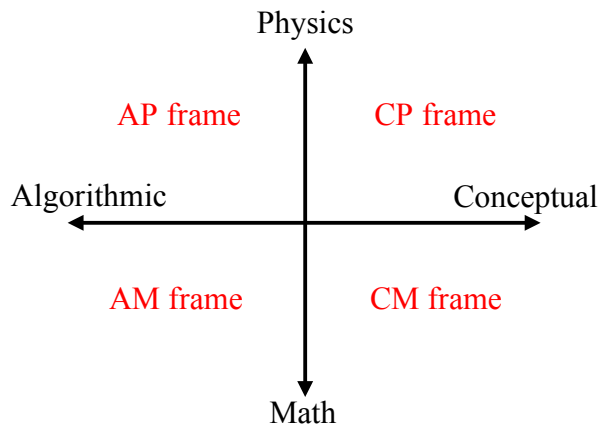

Figure 1. Our two-axis framework. The two axes divide the space into four frames, namely Conceptual Physics (CP), Algorithmic Physics (AP), Algorithmic Math (AM), Conceptual Math (CM)

Conceptual physics (CP) frame: Students are in this frame when discussing conceptual ideas of physics or properties of physics quantities/systems/scenarios. For example, students discuss the physics scenario at hand or make a plan for solving the problem. Students may also determine the direction of vector quantities, and discuss whether a quantity is constant or variable.

Algorithmic physics (AP) frame: Students in this frame combine knowledge of the physical system at hand and appropriate mathematical formulation to set up the mathematical equations for solving the problem. For examples, students may recall an equation or apply a general equation to a specific case or perform dimensional analysis to make sense of an equation or result.

Algorithmic math (AM) frame: Students are in this frame when they perform algorithmic math procedures to obtain a result. For example, students may be computing an integral or solving algebraic equations.

Conceptual math (CM) frame: Students are in this frame if they perceive the computation task at hand as an opportunity to apply mathematical rules and properties to avoid detailed and lengthy manipulation. For example, when students encounter $\int_0^\pi \sin x \cos x\, dx$, they reason that the result is zero based on the orthogonality of the sine and cosine functions instead of actually computing the integral.

## IV. Context and method

The context of this study is an Electromagnetic Fields (E&M) course at a large Midwestern university in the United States. This is an upper-division course covering the first seven chapters of Griffiths' textbook (Griffiths, 1999). We conducted this research in two years of this course (Fall 2013 and 2015) which was taught by the same instructor. Each class had 15 to 20 students divided into groups of three or four. Students are free to choose their groupmates and usually stay in the same group thoughout the course, though occasional students will swap groups.

Classes met for four 50-minute sessions weekly. A typical class started with administrative activities such as homework collection or announcements and a short lecture by the instructor, followed by problem solving. Students discussed and solved problems on a shared whiteboard placed on their table. Some groups sometimes worked independently without much interaction; others were highly collaborative. The instructor frequently visited groups to check students' progress and assist them when necessary.

We collected classroom videos of all 10 groups in the two classes mentioned above for a total of 30 weeks (~110 class days excluding exam days) using a small camera attached to the table of each group, which resulted in ~440 hours of video. The videos were then divided into episodes where each episode covered all activities of a group working on a problem. Therefore, we may have episodes of one group solving several problems and also of one problem being solved by several groups, thus allowing us to look at both the breadth and depth of students' problem solving. The length of each episode ranged from a few minutes to about 30 minutes. We applied the following selection criteria to pare down hundreds of hours of video data to a collection of 50 episodes (or approximately 15 hours of video data) for careful analysis.

- Students are actively engage in verbal or written discourse, and their discussions are clearly audible and their written work is legible. We prefer episodes in which the students completely solve the problems or obtain a reasonable partial solution. The correctness of the solution is inconsequential to our study.

- The problems span a broad range of topics of the E&M course, e.g. the electric/magnetic field in a vacuum or matter, electromagnetic induction, electromagnetic waves, circuits, and co-axial cables. Our preference is to find problems for which there are episodes from more than one group. An example of this type of problem is shown in Figure 2.

**Problem**
A rectangular region has a uniform magnetic field $\vec{B}$ which points out of the page. A rectangular loop of wire is moving inside the region to the right with a constant speed $\vec{v}$. Find the induced electromitive force (emf) in the loop.

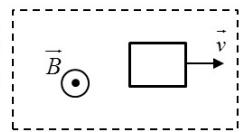

**Solution**
When the whole loop is still inside the magnetic field region, there is no induced emf because the flux through the loop is constant. When the loop is leaving the region, the flux changes and an emf is induced in the loop.
Let x and y be the width inside the magnetic field region and the height of the loop.
The induced electromotive force in the loop is calculated by: $\varepsilon = -\dfrac{d\Phi}{dt}$, in which the flux through the loop is: $\Phi = Bxy$.
Hence: $\varepsilon = -\dfrac{d(Bxy)}{dt} = By\left(-\dfrac{dx}{dt}\right)$
Choose the positive direction to the right, then $v = -\dfrac{dx}{dt}$.
So the induced electromotive force in the loop as it is leaving the region is: $\varepsilon = Byv$

Figure 2. One of the problems selected for our analysis

Within these two selection criteria, the episodes we selected were typical of problem-solving in the course. Other than seeking a breadth of topics across the course, there was no trend to the kind of physics problems or particular physics content that students solved in these episodes. Neither of these two selection criteria are about the framing content of students' work, and the ways that students framed during these problems were typical of their framing throughout the course (with the caveat that the video and audio in these episodes were of sufficient quality for us to analyze their work).

The videos allowed us to investigate students' group work in the most natural setting. We employed an ethnographic approach in analyzing the episodes because it allowed for a closer inspection of student interaction as observed through language, behaviors, and gestures as well as demonstrated an understanding of classroom culture (Paul & Judith, 1998). We watched the videos carefully and wrote narratives about the type, content and duration of each activity that students were engaged in at each specific section of the episode. We defined codes to refer to specific activities and added new codes or merged existing ones as additional episodes were analyzed. Upon completion, we have a codebook consisting of five groups of code, four which correspond to the four frames (conceptual physics, algorithmic physics, algorithmic math, conceptual math) in our framework. The fifth group (O codes) corresponds to the activities or discourse unrelated to any of the four frames, such as students gossiping or the instructor collecting homework.

During an episode, students typically engage in different activities within a frame and across several frames depending on how they are making sense of the activities in which they are engaged. The transition from an activity in one frame to another activity in a different frame is referred to as a frame shift, or in short, a shift. A shift indicates that students are framing a task differently from their previous task. A transition from one activity to another activity within the same frame is not considered a shift since students' framing does not change and students are doing or trying something else within that frame. (Shifts from one O code to another O code are possible, because the O category covers such a wide variety of off-topic activities.)

A frame shift in group problem solving may be initiated by the instructor or members of the group. Sometimes it is a natural shift in the problem solving process which occurs when students finish an activity and progress to the next step in a solution. It is observed in our study that the instructor, through interaction with students group when they are solving problems, affects students' framing to some extent. This is the topic of an ongoing study in our group. In this paper, we focus only on shifts initiated by students and natural shifts occuring during problem solving process.

We classify shifts into three types as indicated in the bulleted list below. Excerpts of students' work from the problem shown in Figure 2 are given as examples.

- **Alpha shift:** A shift to another activity in another frame after students have completed the previous activity. In these shifts, there is no explicit disagreement between group members. For example, students start by recalling the equation for induced electromotive force (AP frame), then discuss how to do the integral and speculate whether they can take $\frac{\partial B}{\partial t}$ out of the integral (AM frame).
- **Beta shift:** A shift similar to the beta shift; however, members of the group disagree or make a bid for a new frame. For example, students reach the point where they attain $\varepsilon = \frac{\partial (B.A)}{\partial t} = \frac{\partial B}{\partial t}(xy)$ and start discussing how to proceed. One student wants to multiply dt over and integrate with respect to dt to get ε(t) (AM frame). However, another student notifies the group that y is constant while x is changing so the area of the loop inside the B-field is changing, so there must be dx/dt (CP frame). In this example, students leave the frame in which they are thinking about math in order to enter the frame where they are considering the properties of physical quantities x and y because a member of the group makes a bid for that change.

- **Gamma shift:** A shift to or from an activity outside of our four quadrants of interest (represented by an O code). For example, students have set up an integral. They proceed to joke with each other for a short while before discussing how to compute the integral.

From an agent-of-change perspective, both alpha and beta shifts are initiated by students, while gamma shifts by be prompted by either instructor or student actions. In this paper, we do not investigate problem solving shifts prompted by the instructor.

## V. Analysis and discussion

In this section, we start with an overall description of possible shifts that groups of students may make during problem solving, including the frames involved, direction of shifts and frequency of shifts between any two frames. Then we examine details regarding the nature of shifts and make sense of them in light of our theoretical framework. The former analysis helps us understand what shifts groups are likely to make and how frequently each shift occurs, while the latter analysis provides insight into how individuals in a group contribute to the framing and the decision of how a frame shifts for the group as a whole.

### A. Within-group frame shifts

Table 2 indicates the starting frame (column) and following frame (row) in every students' transition pair. The number in each cell is the frequency of the shift from the starting frame to the next frame occurence. These shifts occur when students discuss and collaborate within the group during problem solving. The instructor may be present but does not join the students' discussion.

Table 2. Frequency of shifts from Starting frame to Following frame

|  |  | Following frame | | | | | Row total |
|---|---|---|---|---|---|---|---|
|  |  | Other | CP | AP | AM | CM |  |
|  | Other | 1 | 15 | 13 | 18 | 2 | 49 |
| Starting | CP | 13 | 0 | 21 | 13 | 1 | 48 |
| frame | AP | 21 | 16 | 0 | 15 | 0 | 52 |
|  | AM | 14 | 11 | 3 | 0 | 0 | 28 |
|  | CM | 2 | 0 | 1 | 0 | 0 | 3 |
| Column total |  | 50 | 42 | 38 | 46 | 3 | 180 |

We seek to understand the direction of shifts between two frames in each pair by comparing the off-diagonal mirror entries in this table. We notice the large difference between the frequencies of AP-AM and AM-AP shifts, which are 15 and 3, respectively. This is reasonable, because AP-AM shifts most likely occur when students have sufficiently discussed physical principles and recalled necessary physical equations to get started with algorithmic manipulation; while AM-AP shifts occur when students, in the middle of their mathematical manipulation, return to the AP frame to recall additional equations or to rearrange existing equations to support their calculation.

Other pairs of frame do not exhibit large differences between two directions in each pair, which implies that shifts in both directions within a pair are equally likely to happen. CP-AP shifts occur when students have finished exploring the physical scenario and start recalling and applying physical equations to this specific situation. AP-CP shifts are likely to occur when students are recalling equations and mapping back to the physical situation to make sure that they gather the most appropriate equations. With that interpretation, the fact that CP-AP shifts occur at a slightly higher frequency than AP-CP shifts (21 compared to 16) is a reasonable result.

AM-CP and CP-AM shifts also have comparable frequencies (11 and 13, respectively). AM-CP shifts occur when students need to return to examine the physical situation to determine the properties of physical quantities (e.g. whether a quantity is constant or variable, direction of a vector quantity…). CP-AM shifts occur when students have determined the properties of physical quantities and begin solving the math in the equations at hand.

We also notice a large number of shifts into or out of an activity outside of the problem solving process (indicated by Other), such as students sitting quietly, turning in homework or joking. These are natural break points during problem solving in a natural classroom setting.

## *B. Nature of shifts*

The above analysis gives us a general picture of the shifts that occur within groups of students during problem solving. Our focus is now changed to how these shifts occur. Out of 180 shifts presented in Table 2, there are 72 alpha shifts (40%), 9 beta shifts (5%) and 99 gamma shifts (55%).

We first look at gamma shifts because they occur most frequently and involve non-framing activities. As can easily be seen from Table 2, 99 (or 55%) of the shifts within groups are shifts into or out of an Other activity, which are natural break points during problem solving in a real classroom setting. Previous frameworks about students' epistemic framing characterize students' framing at different periods of time into different frames. However, even in each of those periods of time, students may not always be involved in activities of that frame. Students may be discussing or doing unrelated things for such short amounts of time that may be ignored yet lead to inaccurate calculations of the time students spent in each frame (because these activities occur very frequently). Our framework provides a finer analysis of students' framing and therefore allows more accurate calculation of the time students spent in each frame.

Among the remaining shifts, alpha shifts occupy a large portion. They occur when students have finished an activity in one frame and proceed to another activity in a different frame without explicit disagreement from any member of the group. The fact that the vast majority of the shifts between frames (excluding gamma shifts) are alpha shifts doesn't imply that the problems are so easy that students can solve them without any difficulties. Instead, it implies that individuals in a group have similar framing of the problems, thus they can easily reach a concensus about what is taking place and what should occur next while solving problems. This is reasonable since most students in the groups have approximately the same level of physics knowledge and skills, and also because they are generally collegial.

Beta shifts occur when at least one student in the group explicitly raises a question or expresses a disagreement about the activity the group is involved in. The group then has to move to another activity (in another frame) to answer that question or resolve the disagreement before proceeding. Beta shifts also occur when a student makes a bid for a new frame (while the group is in another frame) and that bid is approved and followed by the whole group because they find it reasonable or the frame is more productive. This type of shift rarely occurs (compared to other types of shifts) but it still occupies 5% of the total shifts within groups (or 11% when excluding delta shifts), which implies that it is still a significant phenomenon. More importantly, the investigation of beta shifts allows us to understand how individual students in a group resolve their disagreement or confusion to proceed as a whole group while solving problems.  In contentious groups, we would expect to see many more disagreements; similarly, in groups where one strong student presides over a group of confused peers, we would expect to see many more confusion-driven beta shifts.  The fact that our students do not dissent often doesn't necessarily mean they are too cordial or over-reliant on one student's ideas. In our data, the groups are largely egalitarian: each student contributes ideas, and the marker passed to that student to present his/her ideas to the group. In other words, all students take a turn to "drive the bus" whenever they have an idea.

## VI. Conclusion

Using our two-axis framework, we were able to characterize students' epistemic framing within a group when solving upper-division physics problems in a natural classroom setting. We described how students as a group frame their problem solving activities, what frame shifts they make and how they come to concensus about a frame shift. Characterizing students' frames by quadrant allows us to more finely measure students' framing of mathematics in physics contexts, and asymmetry in frame shifts between quadrants reveals a natural flow to problem solving. Characterizing student frame shifts as alpha or beta allows us to track how often changes in problem solving are driven by consensus or dissent/confusion. The prevalence of gamma shifts suggests that for observational data in a real classroom, students frequently interrupt problem solving to do other things.

These insights may help instructors further understand how group problem solving occurs from a framing perspective and enable them to devise a strategic plan for in-class group work to promote productive framing and effective frame shifts that can be induced by the students themselves. Our study enriches research literature about students' epistemic frames in math in physics context and the dynamics of group framing in problem solving.